\documentclass[showpacs,showkeys,amsmath,amssymb,superscriptaddress,twocolumn]{revtex4-1}
\usepackage[portuges]{babel} 
\usepackage[applemac]{inputenc}
\usepackage[T1]{fontenc} 
\usepackage{makeidx} 
\usepackage{graphicx} 
\usepackage{dcolumn} 
\usepackage{array} 
\usepackage{amssymb} 
\usepackage{amsmath}
\usepackage{textcomp}
\usepackage{rotating}
\usepackage{wasysym}
\usepackage{color}      
\usepackage{multirow}
\usepackage{subfigure}
\usepackage{eucal}
\usepackage{mathrsfs}
\usepackage{units}
\usepackage{rotating}
\usepackage[all]{xy}
\usepackage{float} 
\usepackage{amsmath} 
\usepackage{amsfonts} 
\usepackage{bm} 
\usepackage[amssymb]{SIunits}

\begin{document}

\title{Multifunctional Heusler alloy: experimental evidences of enhanced magnetocaloric properties at room temperature and half-metallicity}
\author{R. J. Caraballo Vivas}
\affiliation{Instituto de F\'{i}sica, Universidade Federal Fluminense, Av. Gal. Milton Tavares de Souza s/n, 24210-346, Niter\'{o}i, RJ, Brazil}
\author{S. S. Pedro}
\affiliation{Instituto de F\'{i}sica, Universidade do Estado do Rio de Janeiro, R. S. Francisco Xavier, 524, Rio de Janeiro, RJ, Brazil}
\author{C.Cruz}
\affiliation{Instituto de F\'{i}sica, Universidade Federal Fluminense, Av. Gal. Milton Tavares de Souza s/n, 24210-346, Niter\'{o}i, RJ, Brazil}
\author{J. C. G. Tedesco}
\affiliation{Instituto de F\'{i}sica, Universidade Federal Fluminense, Av. Gal. Milton Tavares de Souza s/n, 24210-346, Niter\'{o}i, RJ, Brazil}
\author{A. A. Coelho}
\affiliation{Instituto de F\'{i}sica Gleb Wataghin, Universidade Estadual de Campinas - Unicamp, Caixa postal 6165, 13083-859, Campinas, SP, Brazil.}
\author{A. Magnus G. Carvalho}
\affiliation{Laborat\'{o}rio Nacional de Luz S\'{i}ncrotron, CNPEM, Campinas-SP, Brazil.}
\author{D. L. Rocco}
\affiliation{Instituto de F\'{i}sica, Universidade Federal Fluminense, Av. Gal. Milton Tavares de Souza s/n, 24210-346, Niter\'{o}i, RJ, Brazil}
\author{M. S. Reis}
\affiliation{Instituto de F\'{i}sica, Universidade Federal Fluminense, Av. Gal. Milton Tavares de Souza s/n, 24210-346, Niter\'{o}i, RJ, Brazil}

\keywords{Heusler alloys, Magnetocaloric effect, Magnetic refrigeration, Half-metals}


\begin{abstract}
Heusler alloys are widely studied due to their interesting structural and magnetic properties, like magnetic memory shape ability, coupled magneto-structural phase transitions and half-metallicity; ruled, for many cases, by the valence electrons number ($N_v$). The present work focuses on the magnetocaloric potentials of half-metals, exploring the effect of $N_v$ on the magnetic entropy change, preserving half-metallicity. The test bench is the Si-rich side of the half-metallic series Fe$_2$MnSi$_{1-x}$Ga$_x$. From the obtained experimental results it was possible to obtain $|\Delta S|_{max}=\Delta H^{0.8}(\alpha+\beta N_v)$, i.e., the maximum magnetic entropy change depends in a linear fashion on $N_v$, weighted by a power law on the magnetic field change $\Delta H$ ($\alpha$ and $\beta$ are constants experimentally determined). In addition, it was also possible to predict a new multifunctional Heusler alloy, with enhanced magnetocaloric effect,  Curie temperature close to 300 K and half-metallicity.
\end{abstract}

\maketitle

\section{Introduction}

Heusler alloys have been attracted considerable attention due to their several possible applications, such as on spintronics \cite{spin,galanakis}, magneto-optics \cite{Pons200857}, magnetoeletronics \cite{Lielec}, solar thermoeletrics and other technological devices \cite{Graf20111}.  The physical properties for these applications, such as magnetization \cite{galanakis,brown2000magnetization} and the Curie temperature \cite{varaprasad2009,okubo2010magnetic}, can be further optimized managing some parameters, as, for instance, lattice parameter and valence electrons numbers ($N_v$), in which are possible to be ruled by chemical substitution.  An example of the above are alloys that were optimized to exhibit the curious memory shape behavior, defined as the ability of the material to come back to its original shape after deformed by a change in temperature and magnetic field; found, for instance, on Ni$_{45}$Co$_5$Mn$_{36.6}$In$_{13.4}$\cite{kainuma2006}. Also remarkable is the magnetocaloric effect around the magnetic transition temperature due to the occurrence of coupled magneto-structural transitions; found, for instance, on non-stoichometric Ni-Mn-Ga alloys \cite{planes2009,PhysRevB.59.1113}. Other important example is the half-metallicity, in which the alloy presents a gap in the minority band, working therefore as a perfect spin filter, since electrons at the Fermi level are fully polarized. This feature is useful for spintronic purposes and is found, for instance, on Fe$_2$MnSi\cite{Pedro}, Fe$_2$MnP\cite{Kervan} and the high temperature ferromagnets Co$_2$MnSi and Co$_2$MnGa\cite{galanakis}. The aim of this work is thus to predict a multifunctional Heusler alloy, with enhanced magnetocaloric effect at room temperature and half-metallicity. 

More precisely, the magnetocaloric effect (MCE) has been studied by several researchers in order to develop magnetocaloric materials of low cost, good thermal conductivity, low electrical resistivity and mainly maximized magnetocaloric potential. The MCE can be seen from either an adiabatic or isothermal process, both due to a change of the applied magnetic field. From an adiabatic process, the magnetic material changes its temperature; while for an isothermal process it exchanges heat with a thermal reservoir. It is therefore possible to create a thermomagnetic cycle and a magnetic refrigerator based on these processes \cite{BookMario,tishin}. Some compounds that exhibit remarkable MCE potentials are, for instance, manganites \cite{Manga1,Manga2,Andrade,PhysRevB.71.144413}; MnAs-based compounds \cite{MnCuAs,rocconat,Leitao,PhysRevB.77.104439}; Heusler alloys \cite{planes2009,PhysRevB.59.1113}; La-Fe-Si alloys \cite{PhysRevB.67.104416,PhysRevB.65.014410};  intermetallics like RNi$_2$ (R = Nd, Gd, Tb)\cite{plaza}, RCo$_2$ (R = Er, Tb)\cite{Gomes2002870} and PrNi$_{1-x}$Co$_x$ \cite{PhysRevB.79.014428}; and even diamagnetic materials like graphenes\cite{paixao2014oscillating,reis2012oscillating2}.

On the other hand, half-metal materials is one of the key rules to spintronics, since these materials are able to filter majority spins of an incoming non-polarized current. These are therefore useful for tunnel junctions, spin-injection and giant magnetoresistance devices\cite{galanakis}, specially those with high Curie temperature. More precisely, the tunnel magnetoresistance ratio (TMR) become theoretically infinity based on the Julliere's model, for tunnel junctions using half-metals in both electrodes\cite{miyazaki2012physics}. The half-metallicity can be verified from either the theoretical density of states, obtained from first-principle methods, or the total magnetic moment of the compound, that must obey the generalized Slater-Pauling rule $M = (N_v - 24) \mu_B$\cite{galanakis}, where the valence electrons number $N_v$ is written, for our case, as\cite{Pedro}:
\begin{equation}\label{valencias}
N_v = (2 \times N_{\mbox{Fe}}) + N_{\mbox{Mn}} + (1-x) N_{\mbox{Si}}+ xN_{\mbox{Ga}}
\end{equation}
Above, $N_{\mbox{Fe}}= 8$, $N_{\mbox{Mn}}= 7$, $N_{\mbox{Si}}= 4$ and $N_{\mbox{Ga}}= 3$  are the valence electrons for each atom. 

Considering this scenario, our aim is to provide a multifunctional Heusler alloy, with half-metallicity and enhanced magnetocaloric effect, tuning thus this multifunctionallity with the valence electrons number $N_v$. To this purpose, our test bench materials are the Si-rich side of the half-metallic series Fe$_2$MnSi$_{1-x}$Ga$_x$.

\section{Further details on the test bench material}

The magnetic and structural properties of parent Fe$_2$MnSi and Fe$_2$MnGa Heusler alloys have been previously investigated \cite{hiroi2012substitution,kawakami,kudryavtsev}. The former is a well know half-metallic ferromagnet alloy with $T_c$ around 224 K\cite{Pedro} and $Fm\overline{3}m$ spacial group (Cu$_2$MnAl type structure); while Fe$_2$MnGa is also a half-metallic ferromagnet with $T_c$ far above room temperature, around 800 K as previously reported \cite{kudryavtsev,Gasi}, and crystallizes in the $Pm\overline{3}m$ spacial group (Cu$_3$Au type structure). 

\begin{figure}[t!]
\center
\includegraphics[width=0.5\textwidth]{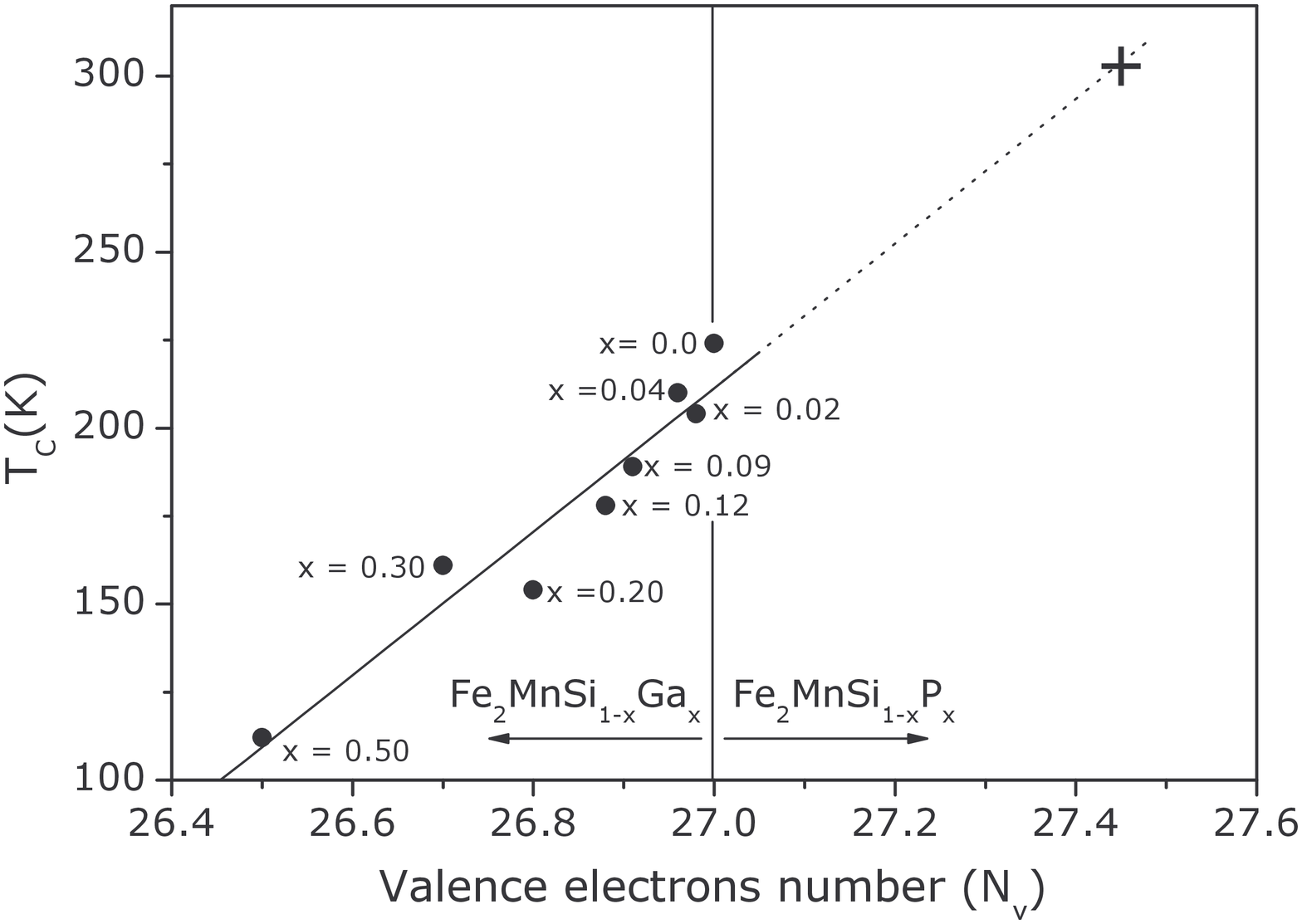}
\caption{Linear behavior of the Curie temperature ($T_c$) as a function of the valence electrons number $N_v$. The  `+' signal marks the necessary $N_v$ value to reach $T_c$ at room temperature, found to be $N_v = 27.44$.}
\label{tc}
\end{figure}

In spite of different crystallographic structures of those parent compounds, it is possible to achieve single phase samples of the series Fe$_2$MnSi$_{1-x}$Ga$_x$; however, only for the Si-rich side up to 50\%; i.e., the crystal structure of Fe$_2$MnSi parent compound supports Ga substitution up to $x=0.50$, with the lattice parameter $a$ increasing by increasing Ga content\cite{Pedro}.  In spite of these finds, the literature has no results on the magnetocaloric effect of these materials, but the Curie temperature of this series was detailed explored by this group and reported in reference \cite{Pedro}. This last was presented as a function of the valence electrons number $N_v$ and an interesting linear behavior was found (see figure \ref{tc}). Thus, since the aim of the present work is to provide a multifunctional Heusler alloy with enhanced magnetocaloric properties ruled by $N_v$ and half-metallicity, from figure \ref{tc} it is straightforward to extrapolate the Curie temperature to 300 K and verify that $N_v = 27.44$ would bring the ferromagnetic transition up to room temperature (a desired feature expected to optimize magnetocaloric materials). Other studies connecting $T_c$ and $N_v$ confirm the linear growth tendency of these quantities for half-metallic Heusler alloys \cite{Graf20111}. 

Thus, as a consequence of the above reported, we must increase the valence electrons number $N_v$ to further optimize the magnetocaloric properties of half-metal Heusler alloys. To achieve this goal, either Si or Ga must be replaced by other element (or elements) that can contribute with more electrons; i.e., those elements belonging to, for instance, group 15 of the periodic table, such as P or As. These elements contribute with 5 electrons and can indeed increase the overall valence electrons number of the system. On the other hand, a substitution by a group 14 element, such as Ge and Sn, does not increase the overall valence electrons number for this series, since these have only 4 valence electrons (the same valence electron number of Si). In fact, Zhang \cite{Zhang2} found values for $T_c$ between only 243 and 260 K in the same structural phase, by replacing Si by Ge in parental Fe$_2$MnSi compound.

From the above, we propose therefore Fe$_2$MnSi$_{0.56}$P$_{0.44}$, since Kervan and Kervan \cite{Kervan} conducted an \textit{ab initio} calculations concerning the Fe$_2$MnP and confirmed the half-metallic features of the systems.

\section{Experimental details}

Polycrystaline ingots of Fe$_{2}$MnSi$_{1-x}$Ga$_{x}$ Heusler alloys were synthesized in arc furnace under Ar atmosphere. The mass of the reactants of high purity were calculated in stoichiometric quantities, with exception of Mn, which was added in excess of 3\% from the stoichiometry to compensate possible losses during the melting process. This additional amount was found using the preparation process described in \cite{Caraballo}.  The ingots were wrapped in tantalum foils, sealed in a quartz tube filled with argon and annealed for 3 days at 1323 K with subsequent quench in water to obtain single phase samples. X-ray powder diffraction data were obtained at room temperature using a Bruker AXS D8 Advance diffractometer with Cu-K$\alpha$ radiation ($\lambda$ = 1.54056 \AA{}) at {\it Laborat\'{o}rio de Difra\c{c}\~{a}o de Raios-X} at UFF and confirmed the single phase formation for all samples. Energy dispersive X-ray spectroscopy (EDS) performed at {\it Laborat\'{o}rio de Caracteriza\c{c}\~{a}o de Materiais} at IF-Sudeste MG was used to obtain the samples composition. The found average values are in very good agreement with the nominal composition. Magnetization data were acquired as a function of temperature and magnetic field  using a commercial Superconducting Quantum Interference Device (SQUID, from Quantum Design$^{\circledR}$) at {\it Laborat\'{o}rio de Baixas Temperaturas} at UNICAMP. Further details about samples preparation and structural characterization can be found in the reference \cite{Pedro}.

\section{Magnetocaloric effect and the valence electrons number}

\begin{figure*}
\center
\includegraphics[width=0.8\textwidth]{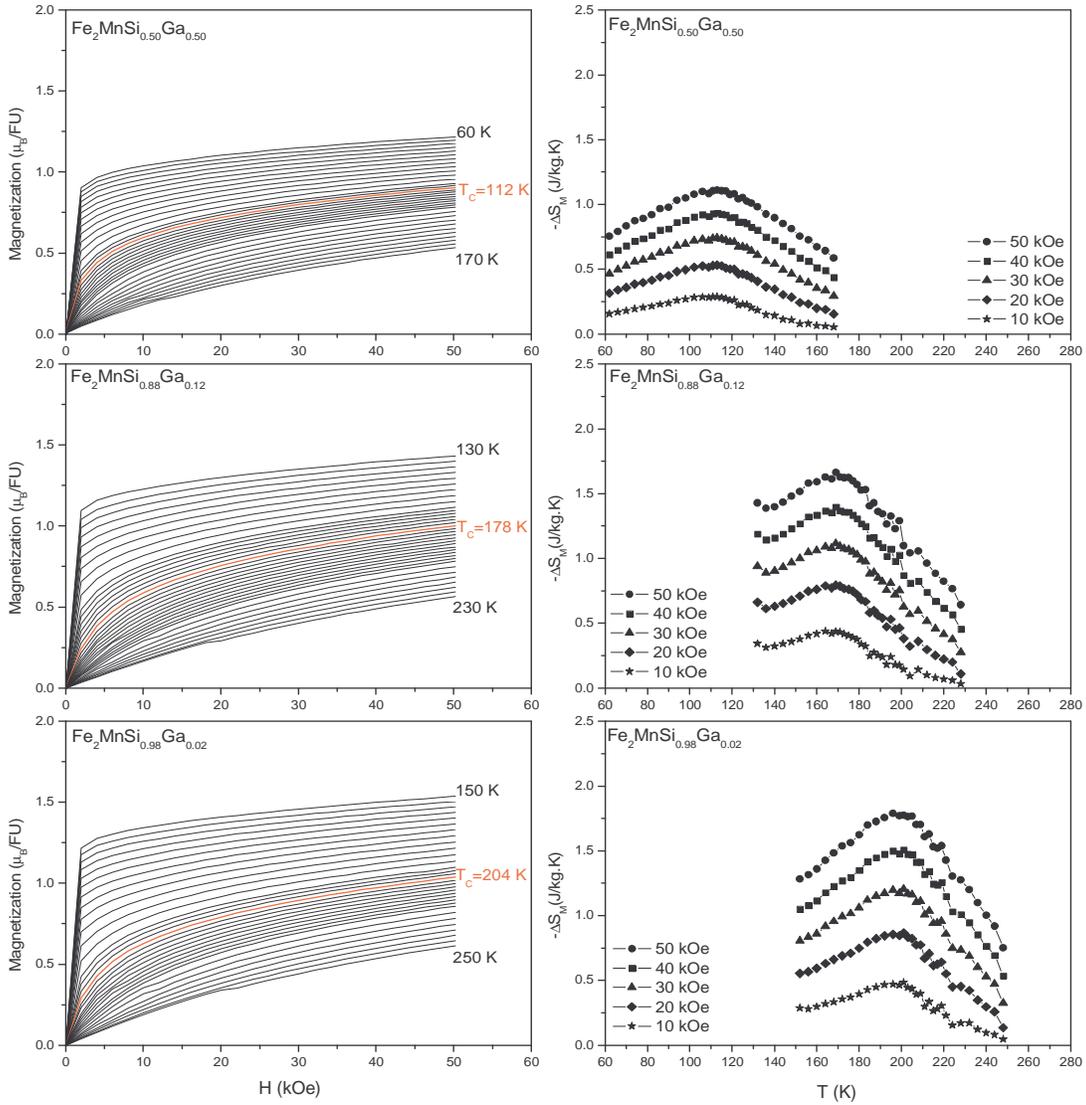}
\caption{(color online) Magnetization isotherms (left) and magnetic entropy changes (right) for the Si-rich side of the Fe$_2$MnSi$_{1-x}$Ga$_x$ series ($x$ = 0.50, 0.12, 0.02).}
\label{mce}
\end{figure*}

The physical quantities that measure the magnetocaloric potential are the magnetic entropy change $\Delta S(T,\Delta H)$ and the adiabatic temperature change $\Delta T(T,\Delta H)$.  The magnetic entropy change is more common to be found in the literature, since it only needs the magnetization map, i.e., $M(T,H)$. Thus, we performed magnetization measurements as a function of magnetic field for several temperatures around the magnetic transition temperature $T_c$ (see figure \ref{mce}-left), for some selected compositions ($x=0.50$, 0.12 and 0.02), chosen  among those available on figure \ref{tc}.

\begin{figure}
\center
\includegraphics[width=0.5\textwidth]{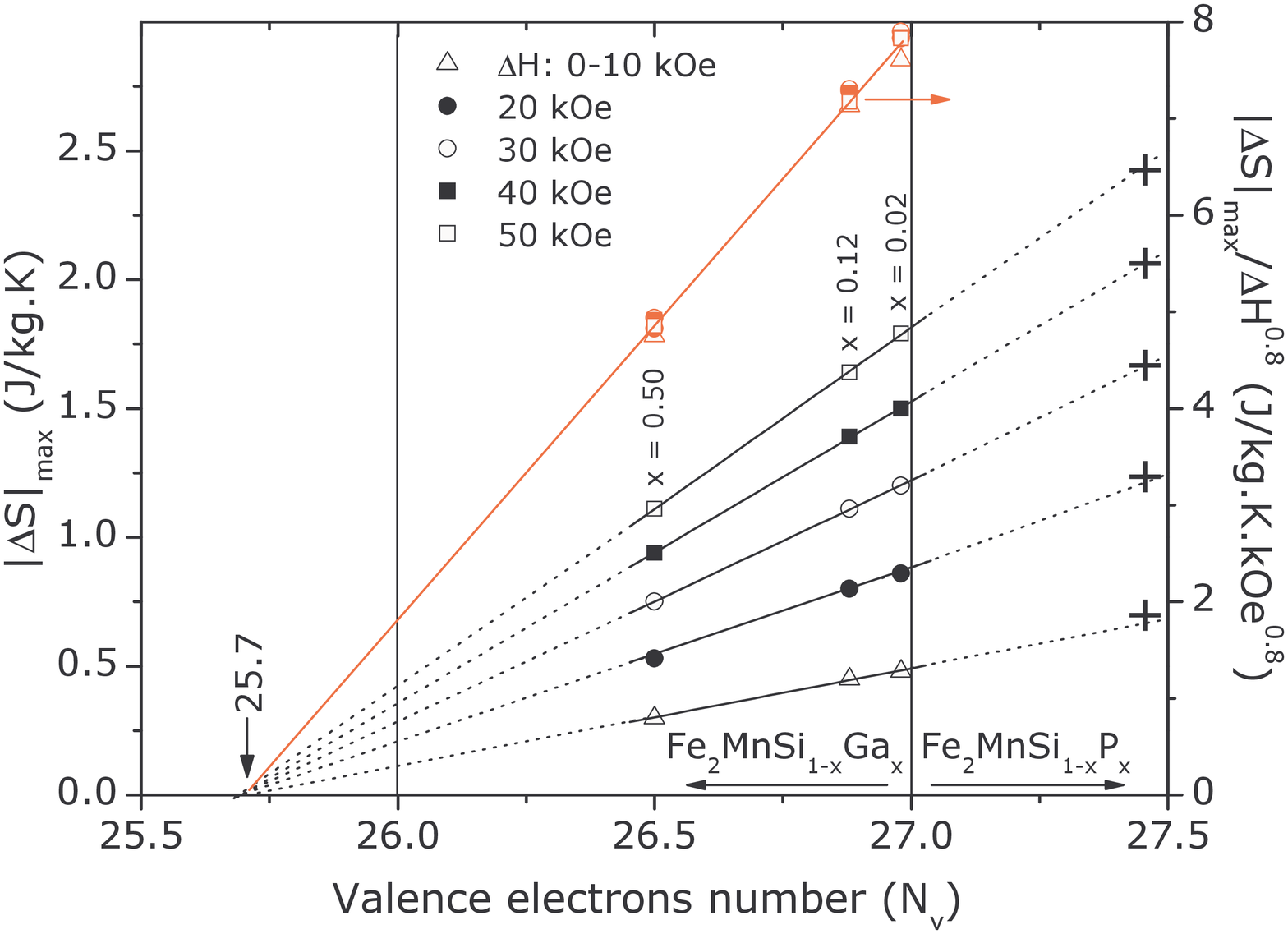}
\caption{(left axis) Maximum magnetic entropy change $|\Delta S|_{max}$ for several values of $\Delta H$ and as a function of $N_v$ for Fe2MnSi$_{1-x}$Ga$_x$ series. The `+' signals mark the possible $|\Delta S|_{max}$ values of the proposed compound with $T_c$ close to room temperature. (right axis-color online) Scaling law to collapse the experimental data, ratifying therefore the validity of equation \ref{finalfwnfew} and the parameters obtained from figure \ref{ab}.}
\label{ds}
\end{figure}

From $M$ \textit{vs.} $H$ isothermal curves it is possible to obtain $\Delta S(T,\Delta H)$ accordingly to:
\begin{equation}
 \Delta S(T,\Delta H)=\int^{H}_{0}\left(\frac{\partial M(T,H)}{\partial T}\right)_{H}dH.
\label{equ1}
\end{equation}
The calculated values of $\Delta S(T,\Delta H)$ are presented on figure \ref{mce}-right for $\Delta H$ = 10, 20, 30, 40 and 50 kOe. A really interesting result was then found by increasing $N_v$ (replacing Ga by Si), in which the maximum magnetic entropy change increases. As mentioned above, the Curie temperature also increases by increasing $N_v$ (see figure \ref{tc}), and therefore a shift towards higher temperatures on the magnetic entropy change peak is observed (and expected). These results are clearly seen on figure \ref{mce}-left and summarized on figure \ref{ds}.

From now on let us then focus on figure \ref{ds}. Note the maximum magnetic entropy change $|\Delta S|_{max}$ has a linear behavior depending on Ga by Si substitution, i.e, depending on $N_v$; and, for $N_v=27.44$, it is expected to achieve 1.2 J/kg.K@20 kOe, that would indeed be comparable to standard metallic Gd (4 J/kg.K@20 kOe). Thus, the Heusler alloy with $N_v=27.44$ would optimize the Curie temperature, shifting $|\Delta S|_{max}$ towards room temperature, and, in addition, enhance the magnetocaloric properties (see figure \ref{ds}).

\begin{figure}
\center
\includegraphics[width=0.5\textwidth]{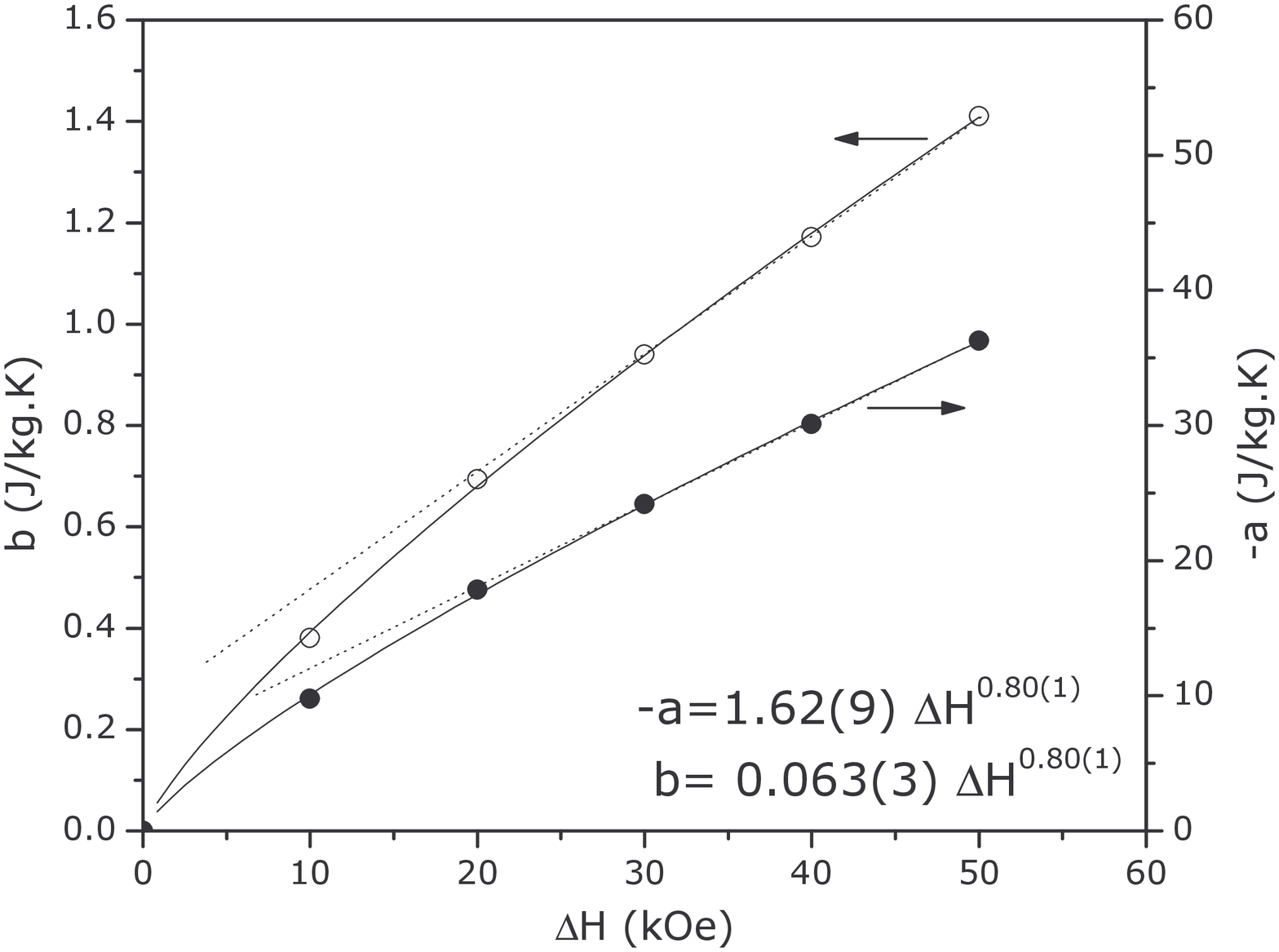}
\caption{Intercept (a) and slope (b) of the linear dependence of $|\Delta S|_{max}$ as a function of $N_v$ (see figure \ref{ds}). These parameters depend on the magnetic field change in a power law fashion $\Delta H^\gamma$ (see equation \ref{abequations}), experimentally found to be $\gamma=0.80(1)$, for both parameters $a$ and $b$. From these dependence the empirical relationship on equation \ref{finalfwnfew} could be obtained. The straight dotted line is a guide to the eyes, to verify the lack of linearity between those parameters and $\Delta H$.}
\label{ab}
\end{figure}

Let us see deeper into this result. Note $|\Delta S|_{max}$ has a linear dependence with $N_v$ for any value of applied field change $\Delta H$, but the slope and intercept parameters of these straight lines are $\Delta H$ dependent. Thus, it is reasonable to propose:
\begin{equation}
|\Delta S|_{max}(N_v,\Delta H)=a(\Delta H)+b(\Delta H) N_v
\end{equation}
Note therefore the value of $N_v$ for which $|\Delta S|_{max}$ goes to zero does not depend on $\Delta H$ and then $a(\Delta H)/b(\Delta H)$ ratio is a constant ($N_v=25.7$), i.e., independent of $\Delta H$. In addition, there is also a boundary condition: $|\Delta S|_{max}(N_v,\Delta H=0)=0$ and thus we can write $a(\Delta H=0)=b(\Delta H=0)=0$. Taking advantage of these ideas, it is reasonable to propose:
\begin{equation}\label{abequations}
a(\Delta H)=\alpha \Delta H^\gamma\;\;\;\text{and}\;\;\;b(\Delta H)=\beta \Delta H^\gamma
\end{equation}
that leads to:
\begin{equation}\label{finalfwnfew}
|\Delta S|_{max}(N_v,\Delta H)=\Delta H^\gamma(\alpha+\beta N_v)
\end{equation}
The above equation satisfies what we are observing. The point now is to obtain the parameters $\alpha$, $\beta$ and $\gamma$ from the experimental results. To go further, we then need $a(\Delta H)$ and $b(\Delta H)$ as a function of $\Delta H$; and these quantities can be easily obtained from the experimental data presented on figure \ref{ds} - and these are shown on figure \ref{ab}.

Equation \ref{abequations} were then fitted to the presented data on figure \ref{ab} and the needed parameters were obtained: $\alpha=-1.62(9)$ J/kg.K.kOe$^\gamma$, $\beta=0.063(3)$ J/kg.K.kOe$^\gamma$ and $\gamma=0.80(1)$. An important point should be emphasized: each fitting has its $\gamma$ exponent free to change and the experimental data lead those two (one from $a$ and the other from $b$), to the same value of 0.80(1). It is an experimental evidence that indeed those straight lines on figure \ref{ds} tend to the same value of $N_v$ for a zero $|\Delta S|_{max}$ value; otherwise we could not factored $\Delta H^{\gamma}$ as it is on equation \ref{finalfwnfew} and, as a consequence, $a(\Delta H)/b(\Delta H)$ would be $\Delta H$ dependent (contrarily to what was observed). To stress this idea, figure \ref{ds}-right axis presents $|\Delta S|_{max}/\Delta H^{0.8}$ and indeed this quantity collapses all of the magnetocaloric data into a single point.

\section{Concluding remarks}

In the present paper we explored the Si-rich side of Fe$_2$MnSi$_{1-x}$Ga$_{x}$ Heusler alloys and concluded that the valence electron number $N_v$ plays an important rule on their Curie temperature and magnetic entropy change. Increasing $N_v$ (equivalente to increase the Si content from Fe$_2$MnSi$_{0.5}$Ga$_{0.5}$ compound), leads to a linear increasing of those both quantities. Our conclusion is that $N_v=27.44$ would bring the Curie temperature of the compound to room temperature, as well as promote an increasing on the maximum magnetic entropy change. From these results we could also propose (based on the experimental data), an empirical linear relationship of the maximum magnetic entropy change $|\Delta S|_{max}$ with $N_v$, weighted by a power law of $\Delta H$, i.e.: $|\Delta S|_{max}=\Delta H^{0.8}(\alpha+\beta N_v)$, where $\alpha$ and $\beta$ are constants experimentally determined. To achieve the goal of the present effort, we also propose to substitute Ga by a group 15 element, like P; and we expect that Fe$_2$MnSi$_{0.56}$P$_{0.44}$ would have their Curie temperatures close to 300 K, with an enhanced magnetocaloric effect. In addition, it is know from reference \cite{Kervan} that Fe$_2$MnP is a half-metal Heusler alloy and therefore we expect that the above proposal leads, in addition to the enhanced MCE properties, a half-metal system with the Curie temperature close to room temperature. Thus, this ideas can indeed lead to new multifunctional material, opening doors for further researches on this topic.

\section{Acknowledgments}

Access to {\it Laborat\'{o}rio de Caracteriza\c{c}\~{a}o de Materiais} at IF-Sudeste MG (Juiz de Fora, Brazil), {\it Laborat\'{o}rio de Difra\c{c}\~ao de Raios-X} at IF - UFF (Niter\'oi, Brazil) and {\it Laborat\'orio de Baixas Temperaturas} at UNICAMP (Campinas, Brazil) are gratefully acknowledged by all authors, in which also acknowledge FAPERJ, FAPESP, CAPES, CNPq and PROPPI-UFF for financial support.


\begin{thebibliography}{40}%
\makeatletter
\providecommand \@ifxundefined [1]{%
 \@ifx{#1\undefined}
}%
\providecommand \@ifnum [1]{%
 \ifnum #1\expandafter \@firstoftwo
 \else \expandafter \@secondoftwo
 \fi
}%
\providecommand \@ifx [1]{%
 \ifx #1\expandafter \@firstoftwo
 \else \expandafter \@secondoftwo
 \fi
}%
\providecommand \natexlab [1]{#1}%
\providecommand \enquote  [1]{``#1''}%
\providecommand \bibnamefont  [1]{#1}%
\providecommand \bibfnamefont [1]{#1}%
\providecommand \citenamefont [1]{#1}%
\providecommand \href@noop [0]{\@secondoftwo}%
\providecommand \href [0]{\begingroup \@sanitize@url \@href}%
\providecommand \@href[1]{\@@startlink{#1}\@@href}%
\providecommand \@@href[1]{\endgroup#1\@@endlink}%
\providecommand \@sanitize@url [0]{\catcode `\\12\catcode `\$12\catcode
  `\&12\catcode `\#12\catcode `\^12\catcode `\_12\catcode `\%12\relax}%
\providecommand \@@startlink[1]{}%
\providecommand \@@endlink[0]{}%
\providecommand \url  [0]{\begingroup\@sanitize@url \@url }%
\providecommand \@url [1]{\endgroup\@href {#1}{\urlprefix }}%
\providecommand \urlprefix  [0]{URL }%
\providecommand \Eprint [0]{\href }%
\providecommand \doibase [0]{http://dx.doi.org/}%
\providecommand \selectlanguage [0]{\@gobble}%
\providecommand \bibinfo  [0]{\@secondoftwo}%
\providecommand \bibfield  [0]{\@secondoftwo}%
\providecommand \translation [1]{[#1]}%
\providecommand \BibitemOpen [0]{}%
\providecommand \bibitemStop [0]{}%
\providecommand \bibitemNoStop [0]{.\EOS\space}%
\providecommand \EOS [0]{\spacefactor3000\relax}%
\providecommand \BibitemShut  [1]{\csname bibitem#1\endcsname}%
\let\auto@bib@innerbib\@empty
\bibitem [{\citenamefont {Wang}\ \emph {et~al.}(2014)\citenamefont {Wang},
  \citenamefont {Meyer}, \citenamefont {Teichert}, \citenamefont {Auge},
  \citenamefont {Rausch}, \citenamefont {Balke}, \citenamefont {Hutten},
  \citenamefont {Fecher},\ and\ \citenamefont {Felser}}]{spin}%
  \BibitemOpen
  \bibfield  {author} {\bibinfo {author} {\bibfnamefont {C.}~\bibnamefont
  {Wang}}, \bibinfo {author} {\bibfnamefont {J.}~\bibnamefont {Meyer}},
  \bibinfo {author} {\bibfnamefont {N.}~\bibnamefont {Teichert}}, \bibinfo
  {author} {\bibfnamefont {A.}~\bibnamefont {Auge}}, \bibinfo {author}
  {\bibfnamefont {E.}~\bibnamefont {Rausch}}, \bibinfo {author} {\bibfnamefont
  {B.}~\bibnamefont {Balke}}, \bibinfo {author} {\bibfnamefont
  {A.}~\bibnamefont {Hutten}}, \bibinfo {author} {\bibfnamefont {G.~H.}\
  \bibnamefont {Fecher}}, \ and\ \bibinfo {author} {\bibfnamefont
  {C.}~\bibnamefont {Felser}},\ }\href {\doibase
  http://dx.doi.org/10.1116/1.4866418} {\bibfield  {journal} {\bibinfo
  {journal} {Journal of Vacuum Science \& Technology B}\ }\textbf {\bibinfo
  {volume} {32}},\ \bibinfo {eid} {020802} (\bibinfo {year}
  {2014})}\BibitemShut {NoStop}%
\bibitem [{\citenamefont {Galanakis}\ and\ \citenamefont
  {H.}(2002)}]{galanakis}%
  \BibitemOpen
  \bibfield  {author} {\bibinfo {author} {\bibfnamefont {I.}~\bibnamefont
  {Galanakis}}\ and\ \bibinfo {author} {\bibfnamefont {D.~P.}\ \bibnamefont
  {H.}},\ }\href {\doibase http://dx.doi.org/0.1103/PhysRevB.66.174429}
  {\bibfield  {journal} {\bibinfo  {journal} {Physical Review B}\ }\textbf
  {\bibinfo {volume} {66}},\ \bibinfo {eid} {-} (\bibinfo {year}
  {2002})}\BibitemShut {NoStop}%
\bibitem [{\citenamefont {Pons}\ \emph {et~al.}(2008)\citenamefont {Pons},
  \citenamefont {Cesari}, \citenamefont {Segu\'i}, \citenamefont {Masdeu},\
  and\ \citenamefont {Santamarta}}]{Pons200857}%
  \BibitemOpen
  \bibfield  {author} {\bibinfo {author} {\bibfnamefont {J.}~\bibnamefont
  {Pons}}, \bibinfo {author} {\bibfnamefont {E.}~\bibnamefont {Cesari}},
  \bibinfo {author} {\bibfnamefont {C.}~\bibnamefont {Segu\'i}}, \bibinfo
  {author} {\bibfnamefont {F.}~\bibnamefont {Masdeu}}, \ and\ \bibinfo {author}
  {\bibfnamefont {R.}~\bibnamefont {Santamarta}},\ }\href {\doibase
  http://dx.doi.org/10.1016/j.msea.2007.02.152} {\bibfield  {journal} {\bibinfo
   {journal} {Materials Science and Engineering: A}\ }\textbf {\bibinfo
  {volume} {481 - 482}},\ \bibinfo {pages} {57 } (\bibinfo {year} {2008})},\
  \bibinfo {note} {proceedings of the 7th European Symposium on Martensitic
  Transformations, \{ESOMAT\} 2006}\BibitemShut {NoStop}%
\bibitem [{\citenamefont {Li}\ \emph {et~al.}(2012)\citenamefont {Li},
  \citenamefont {Liu}, \citenamefont {Lou}, \citenamefont {Beguhn},
  \citenamefont {Wu}, \citenamefont {Qiu}, \citenamefont {Lin}, \citenamefont
  {Cai}, \citenamefont {Hu}, \citenamefont {Xu}, \citenamefont {Duh},\ and\
  \citenamefont {Sun}}]{Lielec}%
  \BibitemOpen
  \bibfield  {author} {\bibinfo {author} {\bibfnamefont {S.}~\bibnamefont
  {Li}}, \bibinfo {author} {\bibfnamefont {M.}~\bibnamefont {Liu}}, \bibinfo
  {author} {\bibfnamefont {J.}~\bibnamefont {Lou}}, \bibinfo {author}
  {\bibfnamefont {S.}~\bibnamefont {Beguhn}}, \bibinfo {author} {\bibfnamefont
  {J.}~\bibnamefont {Wu}}, \bibinfo {author} {\bibfnamefont {J.}~\bibnamefont
  {Qiu}}, \bibinfo {author} {\bibfnamefont {J.}~\bibnamefont {Lin}}, \bibinfo
  {author} {\bibfnamefont {Z.}~\bibnamefont {Cai}}, \bibinfo {author}
  {\bibfnamefont {Y.}~\bibnamefont {Hu}}, \bibinfo {author} {\bibfnamefont
  {F.}~\bibnamefont {Xu}}, \bibinfo {author} {\bibfnamefont {J.-G.}\
  \bibnamefont {Duh}}, \ and\ \bibinfo {author} {\bibfnamefont {N.~X.}\
  \bibnamefont {Sun}},\ }\href@noop {} {\bibfield  {journal} {\bibinfo
  {journal} {Journal of Applied Physics}\ }\textbf {\bibinfo {volume} {111}},\
  \bibinfo {eid} {07} (\bibinfo {year} {2012})}\BibitemShut {NoStop}%
\bibitem [{\citenamefont {Graf}\ \emph {et~al.}(2011)\citenamefont {Graf},
  \citenamefont {Felser},\ and\ \citenamefont {Parkin}}]{Graf20111}%
  \BibitemOpen
  \bibfield  {author} {\bibinfo {author} {\bibfnamefont {T.}~\bibnamefont
  {Graf}}, \bibinfo {author} {\bibfnamefont {C.}~\bibnamefont {Felser}}, \ and\
  \bibinfo {author} {\bibfnamefont {S.~S.}\ \bibnamefont {Parkin}},\ }\href
  {\doibase http://dx.doi.org/10.1016/j.progsolidstchem.2011.02.001} {\bibfield
   {journal} {\bibinfo  {journal} {Progress in Solid State Chemistry}\ }\textbf
  {\bibinfo {volume} {39}},\ \bibinfo {pages} {1 } (\bibinfo {year}
  {2011})}\BibitemShut {NoStop}%
\bibitem [{\citenamefont {Brown}\ \emph {et~al.}(2000)\citenamefont {Brown},
  \citenamefont {Neumann}, \citenamefont {Webster},\ and\ \citenamefont
  {Ziebeck}}]{brown2000magnetization}%
  \BibitemOpen
  \bibfield  {author} {\bibinfo {author} {\bibfnamefont {P.}~\bibnamefont
  {Brown}}, \bibinfo {author} {\bibfnamefont {K.}~\bibnamefont {Neumann}},
  \bibinfo {author} {\bibfnamefont {P.}~\bibnamefont {Webster}}, \ and\
  \bibinfo {author} {\bibfnamefont {K.}~\bibnamefont {Ziebeck}},\ }\href@noop
  {} {\bibfield  {journal} {\bibinfo  {journal} {Journal of Physics: Condensed
  Matter}\ }\textbf {\bibinfo {volume} {12}},\ \bibinfo {pages} {1827}
  (\bibinfo {year} {2000})}\BibitemShut {NoStop}%
\bibitem [{\citenamefont {Varaprasad}\ \emph {et~al.}(2009)\citenamefont
  {Varaprasad}, \citenamefont {Rajanikanth}, \citenamefont {Takahashi},\ and\
  \citenamefont {Hono}}]{varaprasad2009}%
  \BibitemOpen
  \bibfield  {author} {\bibinfo {author} {\bibfnamefont {B.~C.~S.}\
  \bibnamefont {Varaprasad}}, \bibinfo {author} {\bibfnamefont
  {A.}~\bibnamefont {Rajanikanth}}, \bibinfo {author} {\bibfnamefont
  {Y.}~\bibnamefont {Takahashi}}, \ and\ \bibinfo {author} {\bibfnamefont
  {K.}~\bibnamefont {Hono}},\ }\href@noop {} {\bibfield  {journal} {\bibinfo
  {journal} {Acta Materialia}\ }\textbf {\bibinfo {volume} {57}},\ \bibinfo
  {pages} {2702} (\bibinfo {year} {2009})}\BibitemShut {NoStop}%
\bibitem [{\citenamefont {Okubo}\ \emph {et~al.}(2010)\citenamefont {Okubo},
  \citenamefont {Umetsu}, \citenamefont {Kobayashi}, \citenamefont {Kainuma},\
  and\ \citenamefont {Ishida}}]{okubo2010magnetic}%
  \BibitemOpen
  \bibfield  {author} {\bibinfo {author} {\bibfnamefont {A.}~\bibnamefont
  {Okubo}}, \bibinfo {author} {\bibfnamefont {R.}~\bibnamefont {Umetsu}},
  \bibinfo {author} {\bibfnamefont {K.}~\bibnamefont {Kobayashi}}, \bibinfo
  {author} {\bibfnamefont {R.}~\bibnamefont {Kainuma}}, \ and\ \bibinfo
  {author} {\bibfnamefont {K.}~\bibnamefont {Ishida}},\ }\href@noop {}
  {\bibfield  {journal} {\bibinfo  {journal} {Applied Physics Letters}\
  }\textbf {\bibinfo {volume} {96}},\ \bibinfo {pages} {222507} (\bibinfo
  {year} {2010})}\BibitemShut {NoStop}%
\bibitem [{\citenamefont {Kainuma}\ \emph {et~al.}(2006)\citenamefont
  {Kainuma}, \citenamefont {Imano}, \citenamefont {Ito}, \citenamefont {Sutou},
  \citenamefont {Morito}, \citenamefont {Okamoto}, \citenamefont {Kitakami},
  \citenamefont {Oikawa}, \citenamefont {Fujita}, \citenamefont {Kanomata}
  \emph {et~al.}}]{kainuma2006}%
  \BibitemOpen
  \bibfield  {author} {\bibinfo {author} {\bibfnamefont {R.}~\bibnamefont
  {Kainuma}}, \bibinfo {author} {\bibfnamefont {Y.}~\bibnamefont {Imano}},
  \bibinfo {author} {\bibfnamefont {W.}~\bibnamefont {Ito}}, \bibinfo {author}
  {\bibfnamefont {Y.}~\bibnamefont {Sutou}}, \bibinfo {author} {\bibfnamefont
  {H.}~\bibnamefont {Morito}}, \bibinfo {author} {\bibfnamefont
  {S.}~\bibnamefont {Okamoto}}, \bibinfo {author} {\bibfnamefont
  {O.}~\bibnamefont {Kitakami}}, \bibinfo {author} {\bibfnamefont
  {K.}~\bibnamefont {Oikawa}}, \bibinfo {author} {\bibfnamefont
  {A.}~\bibnamefont {Fujita}}, \bibinfo {author} {\bibfnamefont
  {T.}~\bibnamefont {Kanomata}},  \emph {et~al.},\ }\href@noop {} {\bibfield
  {journal} {\bibinfo  {journal} {Nature}\ }\textbf {\bibinfo {volume} {439}},\
  \bibinfo {pages} {957} (\bibinfo {year} {2006})}\BibitemShut {NoStop}%
\bibitem [{\citenamefont {Planes}\ \emph {et~al.}(2009)\citenamefont {Planes},
  \citenamefont {Manosa},\ and\ \citenamefont {Acet}}]{planes2009}%
  \BibitemOpen
  \bibfield  {author} {\bibinfo {author} {\bibfnamefont {A.}~\bibnamefont
  {Planes}}, \bibinfo {author} {\bibfnamefont {L.}~\bibnamefont {Manosa}}, \
  and\ \bibinfo {author} {\bibfnamefont {M.}~\bibnamefont {Acet}},\ }\href@noop
  {} {\bibfield  {journal} {\bibinfo  {journal} {Journal of Physics: Condensed
  Matter}\ }\textbf {\bibinfo {volume} {21}},\ \bibinfo {pages} {233201}
  (\bibinfo {year} {2009})}\BibitemShut {NoStop}%
\bibitem [{\citenamefont {Vasil'ev}\ \emph {et~al.}(1999)\citenamefont
  {Vasil'ev}, \citenamefont {Bozhko}, \citenamefont {Khovailo}, \citenamefont
  {Dikshtein}, \citenamefont {Shavrov}, \citenamefont {Buchelnikov},
  \citenamefont {Matsumoto}, \citenamefont {Suzuki}, \citenamefont {Takagi},\
  and\ \citenamefont {Tani}}]{PhysRevB.59.1113}%
  \BibitemOpen
  \bibfield  {author} {\bibinfo {author} {\bibfnamefont {A.~N.}\ \bibnamefont
  {Vasil'ev}}, \bibinfo {author} {\bibfnamefont {A.~D.}\ \bibnamefont
  {Bozhko}}, \bibinfo {author} {\bibfnamefont {V.~V.}\ \bibnamefont
  {Khovailo}}, \bibinfo {author} {\bibfnamefont {I.~E.}\ \bibnamefont
  {Dikshtein}}, \bibinfo {author} {\bibfnamefont {V.~G.}\ \bibnamefont
  {Shavrov}}, \bibinfo {author} {\bibfnamefont {V.~D.}\ \bibnamefont
  {Buchelnikov}}, \bibinfo {author} {\bibfnamefont {M.}~\bibnamefont
  {Matsumoto}}, \bibinfo {author} {\bibfnamefont {S.}~\bibnamefont {Suzuki}},
  \bibinfo {author} {\bibfnamefont {T.}~\bibnamefont {Takagi}}, \ and\ \bibinfo
  {author} {\bibfnamefont {J.}~\bibnamefont {Tani}},\ }\href {\doibase
  10.1103/PhysRevB.59.1113} {\bibfield  {journal} {\bibinfo  {journal} {Phys.
  Rev. B}\ }\textbf {\bibinfo {volume} {59}},\ \bibinfo {pages} {1113}
  (\bibinfo {year} {1999})}\BibitemShut {NoStop}%
\bibitem [{\citenamefont {Pedro}\ \emph {et~al.}(2015)\citenamefont {Pedro},
  \citenamefont {Caraballo~Vivas}, \citenamefont {Andrade}, \citenamefont
  {Cruz}, \citenamefont {Paix\~ao}, \citenamefont {Contreras}, \citenamefont
  {Costa-Soares}, \citenamefont {Caldeira}, \citenamefont {Coelho},
  \citenamefont {Carvalho}, \citenamefont {Rocco},\ and\ \citenamefont
  {Reis}}]{Pedro}%
  \BibitemOpen
  \bibfield  {author} {\bibinfo {author} {\bibfnamefont {S.~S.}\ \bibnamefont
  {Pedro}}, \bibinfo {author} {\bibfnamefont {R.~J.}\ \bibnamefont
  {Caraballo~Vivas}}, \bibinfo {author} {\bibfnamefont {V.~M.}\ \bibnamefont
  {Andrade}}, \bibinfo {author} {\bibfnamefont {C.}~\bibnamefont {Cruz}},
  \bibinfo {author} {\bibfnamefont {L.~S.}\ \bibnamefont {Paix\~ao}}, \bibinfo
  {author} {\bibfnamefont {C.}~\bibnamefont {Contreras}}, \bibinfo {author}
  {\bibfnamefont {T.}~\bibnamefont {Costa-Soares}}, \bibinfo {author}
  {\bibfnamefont {L.}~\bibnamefont {Caldeira}}, \bibinfo {author}
  {\bibfnamefont {A.~A.}\ \bibnamefont {Coelho}}, \bibinfo {author}
  {\bibfnamefont {A.~M.~G.}\ \bibnamefont {Carvalho}}, \bibinfo {author}
  {\bibfnamefont {D.~L.}\ \bibnamefont {Rocco}}, \ and\ \bibinfo {author}
  {\bibfnamefont {M.~S.}\ \bibnamefont {Reis}},\ }\href {\doibase
  http://dx.doi.org/10.1063/1.4905173} {\bibfield  {journal} {\bibinfo
  {journal} {Journal of Applied Physics}\ }\textbf {\bibinfo {volume} {117}},\
  \bibinfo {eid} {013902} (\bibinfo {year} {2015})}\BibitemShut {NoStop}%
\bibitem [{\citenamefont {Kervan}\ and\ \citenamefont
  {Kervan}(2012{\natexlab{a}})}]{Kervan}%
  \BibitemOpen
  \bibfield  {author} {\bibinfo {author} {\bibfnamefont {N.}~\bibnamefont
  {Kervan}}\ and\ \bibinfo {author} {\bibfnamefont {S.}~\bibnamefont
  {Kervan}},\ }\href {\doibase
  http://dx.doi.org/10.1016/j.intermet.2012.01.030} {\bibfield  {journal}
  {\bibinfo  {journal} {Intermetallics}\ }\textbf {\bibinfo {volume} {24}},\
  \bibinfo {pages} {56 } (\bibinfo {year} {2012}{\natexlab{a}})}\BibitemShut
  {NoStop}%
\bibitem [{\citenamefont {Reis}(2013)}]{BookMario}%
  \BibitemOpen
  \bibfield  {author} {\bibinfo {author} {\bibfnamefont {M.}~\bibnamefont
  {Reis}},\ }\href@noop {} {\emph {\bibinfo {title} {Fundamentals of
  magnetism}}}\ (\bibinfo  {publisher} {Elsevier},\ \bibinfo {year}
  {2013})\BibitemShut {NoStop}%
\bibitem [{\citenamefont {Tishin}\ and\ \citenamefont
  {Spichkin}(2003)}]{tishin}%
  \BibitemOpen
  \bibfield  {author} {\bibinfo {author} {\bibfnamefont {A.~M.}\ \bibnamefont
  {Tishin}}\ and\ \bibinfo {author} {\bibfnamefont {Y.~I.}\ \bibnamefont
  {Spichkin}},\ }\href@noop {} {\emph {\bibinfo {title} {The magnetocaloric
  effect and its applications}}}\ (\bibinfo  {publisher} {CRC Press},\ \bibinfo
  {year} {2003})\BibitemShut {NoStop}%
\bibitem [{\citenamefont {Rocco}\ \emph {et~al.}(2013)\citenamefont {Rocco},
  \citenamefont {Coelho}, \citenamefont {Gama},\ and\ \citenamefont
  {Santos}}]{Manga1}%
  \BibitemOpen
  \bibfield  {author} {\bibinfo {author} {\bibfnamefont {D.~L.}\ \bibnamefont
  {Rocco}}, \bibinfo {author} {\bibfnamefont {A.~A.}\ \bibnamefont {Coelho}},
  \bibinfo {author} {\bibfnamefont {S.}~\bibnamefont {Gama}}, \ and\ \bibinfo
  {author} {\bibfnamefont {M.~d.~C.}\ \bibnamefont {Santos}},\ }\href {\doibase
  http://dx.doi.org/10.1063/1.4795769} {\bibfield  {journal} {\bibinfo
  {journal} {Journal of Applied Physics}\ }\textbf {\bibinfo {volume} {113}},\
  \bibinfo {eid} {113907} (\bibinfo {year} {2013})}\BibitemShut {NoStop}%
\bibitem [{\citenamefont {Rocco}\ \emph {et~al.}(2005)\citenamefont {Rocco},
  \citenamefont {Silva}, \citenamefont {Carvalho}, \citenamefont {Coelho},
  \citenamefont {Andreeta},\ and\ \citenamefont {Gama}}]{Manga2}%
  \BibitemOpen
  \bibfield  {author} {\bibinfo {author} {\bibfnamefont {D.~L.}\ \bibnamefont
  {Rocco}}, \bibinfo {author} {\bibfnamefont {R.~A.}\ \bibnamefont {Silva}},
  \bibinfo {author} {\bibfnamefont {A.~M.~G.}\ \bibnamefont {Carvalho}},
  \bibinfo {author} {\bibfnamefont {A.~A.}\ \bibnamefont {Coelho}}, \bibinfo
  {author} {\bibfnamefont {J.~P.}\ \bibnamefont {Andreeta}}, \ and\ \bibinfo
  {author} {\bibfnamefont {S.}~\bibnamefont {Gama}},\ }\href@noop {} {\bibfield
   {journal} {\bibinfo  {journal} {Journal of Applied Physics}\ }\textbf
  {\bibinfo {volume} {97}},\ \bibinfo {eid} {10} (\bibinfo {year}
  {2005})}\BibitemShut {NoStop}%
\bibitem [{\citenamefont {Andrade}\ \emph {et~al.}(2014)\citenamefont
  {Andrade}, \citenamefont {Caraballo-Vivas}, \citenamefont {Costas-Soares},
  \citenamefont {Pedro}, \citenamefont {Rocco}, \citenamefont {Reis},
  \citenamefont {Campos},\ and\ \citenamefont {Coelho}}]{Andrade}%
  \BibitemOpen
  \bibfield  {author} {\bibinfo {author} {\bibfnamefont {V.}~\bibnamefont
  {Andrade}}, \bibinfo {author} {\bibfnamefont {R.}~\bibnamefont
  {Caraballo-Vivas}}, \bibinfo {author} {\bibfnamefont {T.}~\bibnamefont
  {Costas-Soares}}, \bibinfo {author} {\bibfnamefont {S.}~\bibnamefont
  {Pedro}}, \bibinfo {author} {\bibfnamefont {D.}~\bibnamefont {Rocco}},
  \bibinfo {author} {\bibfnamefont {M.}~\bibnamefont {Reis}}, \bibinfo {author}
  {\bibfnamefont {A.}~\bibnamefont {Campos}}, \ and\ \bibinfo {author}
  {\bibfnamefont {A.}~\bibnamefont {Coelho}},\ }\href {\doibase
  http://dx.doi.org/10.1016/j.jssc.2014.07.013} {\bibfield  {journal} {\bibinfo
   {journal} {Journal of Solid State Chemistry}\ }\textbf {\bibinfo {volume}
  {219}},\ \bibinfo {pages} {87 } (\bibinfo {year} {2014})}\BibitemShut
  {NoStop}%
\bibitem [{\citenamefont {Reis}\ \emph {et~al.}(2005)\citenamefont {Reis},
  \citenamefont {Amaral}, \citenamefont {Ara\'ujo}, \citenamefont {Tavares},
  \citenamefont {Gomes},\ and\ \citenamefont {Oliveira}}]{PhysRevB.71.144413}%
  \BibitemOpen
  \bibfield  {author} {\bibinfo {author} {\bibfnamefont {M.~S.}\ \bibnamefont
  {Reis}}, \bibinfo {author} {\bibfnamefont {V.~S.}\ \bibnamefont {Amaral}},
  \bibinfo {author} {\bibfnamefont {J.~P.}\ \bibnamefont {Ara\'ujo}}, \bibinfo
  {author} {\bibfnamefont {P.~B.}\ \bibnamefont {Tavares}}, \bibinfo {author}
  {\bibfnamefont {A.~M.}\ \bibnamefont {Gomes}}, \ and\ \bibinfo {author}
  {\bibfnamefont {I.~S.}\ \bibnamefont {Oliveira}},\ }\href {\doibase
  10.1103/PhysRevB.71.144413} {\bibfield  {journal} {\bibinfo  {journal} {Phys.
  Rev. B}\ }\textbf {\bibinfo {volume} {71}},\ \bibinfo {pages} {144413}
  (\bibinfo {year} {2005})}\BibitemShut {NoStop}%
\bibitem [{\citenamefont {Rocco}\ \emph {et~al.}(2007)\citenamefont {Rocco},
  \citenamefont {de~Campos}, \citenamefont {Carvalho}, \citenamefont {Caron},
  \citenamefont {Coelho}, \citenamefont {Gama}, \citenamefont {Gandra},
  \citenamefont {dos Santos}, \citenamefont {Cardoso}, \citenamefont {von
  Ranke},\ and\ \citenamefont {de~Oliveira}}]{MnCuAs}%
  \BibitemOpen
  \bibfield  {author} {\bibinfo {author} {\bibfnamefont {D.~L.}\ \bibnamefont
  {Rocco}}, \bibinfo {author} {\bibfnamefont {A.}~\bibnamefont {de~Campos}},
  \bibinfo {author} {\bibfnamefont {A.~M.~G.}\ \bibnamefont {Carvalho}},
  \bibinfo {author} {\bibfnamefont {L.}~\bibnamefont {Caron}}, \bibinfo
  {author} {\bibfnamefont {A.~A.}\ \bibnamefont {Coelho}}, \bibinfo {author}
  {\bibfnamefont {S.}~\bibnamefont {Gama}}, \bibinfo {author} {\bibfnamefont
  {F.~C.~G.}\ \bibnamefont {Gandra}}, \bibinfo {author} {\bibfnamefont {A.~O.}\
  \bibnamefont {dos Santos}}, \bibinfo {author} {\bibfnamefont {L.~P.}\
  \bibnamefont {Cardoso}}, \bibinfo {author} {\bibfnamefont {P.~J.}\
  \bibnamefont {von Ranke}}, \ and\ \bibinfo {author} {\bibfnamefont {N.~A.}\
  \bibnamefont {de~Oliveira}},\ }\href {\doibase
  http://dx.doi.org/10.1063/1.2746074} {\bibfield  {journal} {\bibinfo
  {journal} {Applied Physics Letters}\ }\textbf {\bibinfo {volume} {90}},\
  \bibinfo {eid} {242507} (\bibinfo {year} {2007})}\BibitemShut {NoStop}%
\bibitem [{\citenamefont {De~Campos}\ \emph {et~al.}(2006)\citenamefont
  {De~Campos}, \citenamefont {Rocco}, \citenamefont {Carvalho}, \citenamefont
  {Caron}, \citenamefont {Coelho}, \citenamefont {Gama}, \citenamefont
  {Da~Silva}, \citenamefont {Gandra}, \citenamefont {Dos~Santos}, \citenamefont
  {Cardoso} \emph {et~al.}}]{rocconat}%
  \BibitemOpen
  \bibfield  {author} {\bibinfo {author} {\bibfnamefont {A.}~\bibnamefont
  {De~Campos}}, \bibinfo {author} {\bibfnamefont {D.~L.}\ \bibnamefont
  {Rocco}}, \bibinfo {author} {\bibfnamefont {A.~M.~G.}\ \bibnamefont
  {Carvalho}}, \bibinfo {author} {\bibfnamefont {L.}~\bibnamefont {Caron}},
  \bibinfo {author} {\bibfnamefont {A.~A.}\ \bibnamefont {Coelho}}, \bibinfo
  {author} {\bibfnamefont {S.}~\bibnamefont {Gama}}, \bibinfo {author}
  {\bibfnamefont {L.~M.}\ \bibnamefont {Da~Silva}}, \bibinfo {author}
  {\bibfnamefont {F.~C.}\ \bibnamefont {Gandra}}, \bibinfo {author}
  {\bibfnamefont {A.~O.}\ \bibnamefont {Dos~Santos}}, \bibinfo {author}
  {\bibfnamefont {L.~P.}\ \bibnamefont {Cardoso}},  \emph {et~al.},\
  }\href@noop {} {\bibfield  {journal} {\bibinfo  {journal} {Nature materials}\
  }\textbf {\bibinfo {volume} {5}},\ \bibinfo {pages} {802} (\bibinfo {year}
  {2006})}\BibitemShut {NoStop}%
\bibitem [{\citenamefont {Lei\~tao}\ \emph {et~al.}(2008)\citenamefont
  {Lei\~tao}, \citenamefont {Rocco}, \citenamefont {Sequeira~Amaral},
  \citenamefont {Reis}, \citenamefont {Amaral}, \citenamefont {Fernandes},
  \citenamefont {Martins},\ and\ \citenamefont {Tavares}}]{Leitao}%
  \BibitemOpen
  \bibfield  {author} {\bibinfo {author} {\bibfnamefont {J.}~\bibnamefont
  {Lei\~tao}}, \bibinfo {author} {\bibfnamefont {D.}~\bibnamefont {Rocco}},
  \bibinfo {author} {\bibfnamefont {J.}~\bibnamefont {Sequeira~Amaral}},
  \bibinfo {author} {\bibfnamefont {M.}~\bibnamefont {Reis}}, \bibinfo {author}
  {\bibfnamefont {V.}~\bibnamefont {Amaral}}, \bibinfo {author} {\bibfnamefont
  {R.}~\bibnamefont {Fernandes}}, \bibinfo {author} {\bibfnamefont
  {N.}~\bibnamefont {Martins}}, \ and\ \bibinfo {author} {\bibfnamefont
  {P.}~\bibnamefont {Tavares}},\ }\href {\doibase 10.1109/TMAG.2008.2002794}
  {\bibfield  {journal} {\bibinfo  {journal} {Magnetics, IEEE Transactions on}\
  }\textbf {\bibinfo {volume} {44}},\ \bibinfo {pages} {3036} (\bibinfo {year}
  {2008})}\BibitemShut {NoStop}%
\bibitem [{\citenamefont {Reis}\ \emph {et~al.}(2008)\citenamefont {Reis},
  \citenamefont {Rubinger}, \citenamefont {Sobolev}, \citenamefont {Valente},
  \citenamefont {Yamada}, \citenamefont {Sato}, \citenamefont {Todate},
  \citenamefont {Bouravleuv}, \citenamefont {von Ranke},\ and\ \citenamefont
  {Gama}}]{PhysRevB.77.104439}%
  \BibitemOpen
  \bibfield  {author} {\bibinfo {author} {\bibfnamefont {M.~S.}\ \bibnamefont
  {Reis}}, \bibinfo {author} {\bibfnamefont {R.~M.}\ \bibnamefont {Rubinger}},
  \bibinfo {author} {\bibfnamefont {N.~A.}\ \bibnamefont {Sobolev}}, \bibinfo
  {author} {\bibfnamefont {M.~A.}\ \bibnamefont {Valente}}, \bibinfo {author}
  {\bibfnamefont {K.}~\bibnamefont {Yamada}}, \bibinfo {author} {\bibfnamefont
  {K.}~\bibnamefont {Sato}}, \bibinfo {author} {\bibfnamefont {Y.}~\bibnamefont
  {Todate}}, \bibinfo {author} {\bibfnamefont {A.}~\bibnamefont {Bouravleuv}},
  \bibinfo {author} {\bibfnamefont {P.~J.}\ \bibnamefont {von Ranke}}, \ and\
  \bibinfo {author} {\bibfnamefont {S.}~\bibnamefont {Gama}},\ }\href {\doibase
  10.1103/PhysRevB.77.104439} {\bibfield  {journal} {\bibinfo  {journal} {Phys.
  Rev. B}\ }\textbf {\bibinfo {volume} {77}},\ \bibinfo {pages} {104439}
  (\bibinfo {year} {2008})}\BibitemShut {NoStop}%
\bibitem [{\citenamefont {Fujita}\ \emph {et~al.}(2003)\citenamefont {Fujita},
  \citenamefont {Fujieda}, \citenamefont {Hasegawa},\ and\ \citenamefont
  {Fukamichi}}]{PhysRevB.67.104416}%
  \BibitemOpen
  \bibfield  {author} {\bibinfo {author} {\bibfnamefont {A.}~\bibnamefont
  {Fujita}}, \bibinfo {author} {\bibfnamefont {S.}~\bibnamefont {Fujieda}},
  \bibinfo {author} {\bibfnamefont {Y.}~\bibnamefont {Hasegawa}}, \ and\
  \bibinfo {author} {\bibfnamefont {K.}~\bibnamefont {Fukamichi}},\ }\href
  {\doibase 10.1103/PhysRevB.67.104416} {\bibfield  {journal} {\bibinfo
  {journal} {Phys. Rev. B}\ }\textbf {\bibinfo {volume} {67}},\ \bibinfo
  {pages} {104416} (\bibinfo {year} {2003})}\BibitemShut {NoStop}%
\bibitem [{\citenamefont {Fujita}\ \emph {et~al.}(2001)\citenamefont {Fujita},
  \citenamefont {Fujieda}, \citenamefont {Fukamichi}, \citenamefont
  {Mitamura},\ and\ \citenamefont {Goto}}]{PhysRevB.65.014410}%
  \BibitemOpen
  \bibfield  {author} {\bibinfo {author} {\bibfnamefont {A.}~\bibnamefont
  {Fujita}}, \bibinfo {author} {\bibfnamefont {S.}~\bibnamefont {Fujieda}},
  \bibinfo {author} {\bibfnamefont {K.}~\bibnamefont {Fukamichi}}, \bibinfo
  {author} {\bibfnamefont {H.}~\bibnamefont {Mitamura}}, \ and\ \bibinfo
  {author} {\bibfnamefont {T.}~\bibnamefont {Goto}},\ }\href {\doibase
  10.1103/PhysRevB.65.014410} {\bibfield  {journal} {\bibinfo  {journal} {Phys.
  Rev. B}\ }\textbf {\bibinfo {volume} {65}},\ \bibinfo {pages} {014410}
  (\bibinfo {year} {2001})}\BibitemShut {NoStop}%
\bibitem [{\citenamefont {Plaza}\ \emph {et~al.}(2009)\citenamefont {Plaza},
  \citenamefont {de~Sousa}, \citenamefont {von Ranke}, \citenamefont {Gomes},
  \citenamefont {Rocco}, \citenamefont {Leit\~ao},\ and\ \citenamefont
  {Reis}}]{plaza}%
  \BibitemOpen
  \bibfield  {author} {\bibinfo {author} {\bibfnamefont {E.~J.~R.}\
  \bibnamefont {Plaza}}, \bibinfo {author} {\bibfnamefont {V.~S.~R.}\
  \bibnamefont {de~Sousa}}, \bibinfo {author} {\bibfnamefont {P.~J.}\
  \bibnamefont {von Ranke}}, \bibinfo {author} {\bibfnamefont {A.~M.}\
  \bibnamefont {Gomes}}, \bibinfo {author} {\bibfnamefont {D.~L.}\ \bibnamefont
  {Rocco}}, \bibinfo {author} {\bibfnamefont {J.~V.}\ \bibnamefont {Leit\~ao}},
  \ and\ \bibinfo {author} {\bibfnamefont {M.~S.}\ \bibnamefont {Reis}},\
  }\href {\doibase http://dx.doi.org/10.1063/1.3054178} {\bibfield  {journal}
  {\bibinfo  {journal} {Journal of Applied Physics}\ }\textbf {\bibinfo
  {volume} {105}},\ \bibinfo {eid} {013903} (\bibinfo {year}
  {2009})}\BibitemShut {NoStop}%
\bibitem [{\citenamefont {Gomes}\ \emph {et~al.}(2002)\citenamefont {Gomes},
  \citenamefont {Reis}, \citenamefont {Oliveira}, \citenamefont {aes},\ and\
  \citenamefont {Takeuchi}}]{Gomes2002870}%
  \BibitemOpen
  \bibfield  {author} {\bibinfo {author} {\bibfnamefont {A.}~\bibnamefont
  {Gomes}}, \bibinfo {author} {\bibfnamefont {M.}~\bibnamefont {Reis}},
  \bibinfo {author} {\bibfnamefont {I.}~\bibnamefont {Oliveira}}, \bibinfo
  {author} {\bibfnamefont {A.~G.}\ \bibnamefont {aes}}, \ and\ \bibinfo
  {author} {\bibfnamefont {A.}~\bibnamefont {Takeuchi}},\ }\href {\doibase
  http://dx.doi.org/10.1016/S0304-8853(01)01327-0} {\bibfield  {journal}
  {\bibinfo  {journal} {Journal of Magnetism and Magnetic Materials}\ }\textbf
  {\bibinfo {volume} {242}},\ \bibinfo {pages} {870 } (\bibinfo {year}
  {2002})}\BibitemShut {NoStop}%
\bibitem [{\citenamefont {Rocco}\ \emph {et~al.}(2009)\citenamefont {Rocco},
  \citenamefont {Amaral}, \citenamefont {Leit\~ao}, \citenamefont {Amaral},
  \citenamefont {Reis}, \citenamefont {Fernandes}, \citenamefont {Pereira},
  \citenamefont {Ara\'ujo}, \citenamefont {Martins}, \citenamefont {Tavares},\
  and\ \citenamefont {Coelho}}]{PhysRevB.79.014428}%
  \BibitemOpen
  \bibfield  {author} {\bibinfo {author} {\bibfnamefont {D.~L.}\ \bibnamefont
  {Rocco}}, \bibinfo {author} {\bibfnamefont {J.~S.}\ \bibnamefont {Amaral}},
  \bibinfo {author} {\bibfnamefont {J.~V.}\ \bibnamefont {Leit\~ao}}, \bibinfo
  {author} {\bibfnamefont {V.~S.}\ \bibnamefont {Amaral}}, \bibinfo {author}
  {\bibfnamefont {M.~S.}\ \bibnamefont {Reis}}, \bibinfo {author}
  {\bibfnamefont {R.~P.}\ \bibnamefont {Fernandes}}, \bibinfo {author}
  {\bibfnamefont {A.~M.}\ \bibnamefont {Pereira}}, \bibinfo {author}
  {\bibfnamefont {J.~P.}\ \bibnamefont {Ara\'ujo}}, \bibinfo {author}
  {\bibfnamefont {N.~V.}\ \bibnamefont {Martins}}, \bibinfo {author}
  {\bibfnamefont {P.~B.}\ \bibnamefont {Tavares}}, \ and\ \bibinfo {author}
  {\bibfnamefont {A.~A.}\ \bibnamefont {Coelho}},\ }\href {\doibase
  10.1103/PhysRevB.79.014428} {\bibfield  {journal} {\bibinfo  {journal} {Phys.
  Rev. B}\ }\textbf {\bibinfo {volume} {79}},\ \bibinfo {pages} {014428}
  (\bibinfo {year} {2009})}\BibitemShut {NoStop}%
\bibitem [{\citenamefont {Paix{\~a}o}\ \emph {et~al.}(2014)\citenamefont
  {Paix{\~a}o}, \citenamefont {Alisultanov},\ and\ \citenamefont
  {Reis}}]{paixao2014oscillating}%
  \BibitemOpen
  \bibfield  {author} {\bibinfo {author} {\bibfnamefont {L.}~\bibnamefont
  {Paix{\~a}o}}, \bibinfo {author} {\bibfnamefont {Z.}~\bibnamefont
  {Alisultanov}}, \ and\ \bibinfo {author} {\bibfnamefont {M.}~\bibnamefont
  {Reis}},\ }\href@noop {} {\bibfield  {journal} {\bibinfo  {journal} {Journal
  of Magnetism and Magnetic Materials}\ }\textbf {\bibinfo {volume} {368}},\
  \bibinfo {pages} {374} (\bibinfo {year} {2014})}\BibitemShut {NoStop}%
\bibitem [{\citenamefont {Reis}(2012)}]{reis2012oscillating2}%
  \BibitemOpen
  \bibfield  {author} {\bibinfo {author} {\bibfnamefont {M.}~\bibnamefont
  {Reis}},\ }\href@noop {} {\bibfield  {journal} {\bibinfo  {journal} {Applied
  Physics Letters}\ }\textbf {\bibinfo {volume} {101}},\ \bibinfo {pages}
  {222405} (\bibinfo {year} {2012})}\BibitemShut {NoStop}%
\bibitem [{\citenamefont {Miyazaki}\ and\ \citenamefont
  {Jin}(2012)}]{miyazaki2012physics}%
  \BibitemOpen
  \bibfield  {author} {\bibinfo {author} {\bibfnamefont {T.}~\bibnamefont
  {Miyazaki}}\ and\ \bibinfo {author} {\bibfnamefont {H.}~\bibnamefont {Jin}},\
  }\href@noop {} {\emph {\bibinfo {title} {The physics of ferromagnetism}}},\
  Vol.\ \bibinfo {volume} {158}\ (\bibinfo  {publisher} {Springer Science \&
  Business Media},\ \bibinfo {year} {2012})\BibitemShut {NoStop}%
\bibitem [{\citenamefont {Kervan}\ and\ \citenamefont
  {Kervan}(2012{\natexlab{b}})}]{kervan2012half}%
  \BibitemOpen
  \bibfield  {author} {\bibinfo {author} {\bibfnamefont {N.}~\bibnamefont
  {Kervan}}\ and\ \bibinfo {author} {\bibfnamefont {S.}~\bibnamefont
  {Kervan}},\ }\href@noop {} {\bibfield  {journal} {\bibinfo  {journal}
  {Intermetallics}\ }\textbf {\bibinfo {volume} {24}},\ \bibinfo {pages} {56}
  (\bibinfo {year} {2012}{\natexlab{b}})}\BibitemShut {NoStop}%
\bibitem [{\citenamefont {Hiroi}\ \emph {et~al.}(2012)\citenamefont {Hiroi},
  \citenamefont {Yano}, \citenamefont {Sezaki}, \citenamefont {Shigeta},
  \citenamefont {Ito}, \citenamefont {Manaka},\ and\ \citenamefont
  {Terada}}]{hiroi2012substitution}%
  \BibitemOpen
  \bibfield  {author} {\bibinfo {author} {\bibfnamefont {M.}~\bibnamefont
  {Hiroi}}, \bibinfo {author} {\bibfnamefont {I.}~\bibnamefont {Yano}},
  \bibinfo {author} {\bibfnamefont {K.}~\bibnamefont {Sezaki}}, \bibinfo
  {author} {\bibfnamefont {I.}~\bibnamefont {Shigeta}}, \bibinfo {author}
  {\bibfnamefont {M.}~\bibnamefont {Ito}}, \bibinfo {author} {\bibfnamefont
  {H.}~\bibnamefont {Manaka}}, \ and\ \bibinfo {author} {\bibfnamefont
  {N.}~\bibnamefont {Terada}},\ }in\ \href@noop {} {\emph {\bibinfo {booktitle}
  {Journal of Physics: Conference Series}}},\ Vol.\ \bibinfo {volume} {400}\
  (\bibinfo {organization} {IOP Publishing},\ \bibinfo {year} {2012})\ p.\
  \bibinfo {pages} {032021}\BibitemShut {NoStop}%
\bibitem [{\citenamefont {Nagano}\ \emph {et~al.}(1995)\citenamefont {Nagano},
  \citenamefont {Uwanuyu},\ and\ \citenamefont {Kawakami}}]{kawakami}%
  \BibitemOpen
  \bibfield  {author} {\bibinfo {author} {\bibfnamefont {T.}~\bibnamefont
  {Nagano}}, \bibinfo {author} {\bibfnamefont {S.}~\bibnamefont {Uwanuyu}}, \
  and\ \bibinfo {author} {\bibfnamefont {M.}~\bibnamefont {Kawakami}},\
  }\href@noop {} {\bibfield  {journal} {\bibinfo  {journal} {Journal of
  magnetism and magnetic materials}\ }\textbf {\bibinfo {volume} {140}},\
  \bibinfo {pages} {123} (\bibinfo {year} {1995})}\BibitemShut {NoStop}%
\bibitem [{\citenamefont {Kudryavtsev}\ \emph {et~al.}(2012)\citenamefont
  {Kudryavtsev}, \citenamefont {Uvarov}, \citenamefont {Iermolenko},
  \citenamefont {Glavatskyy},\ and\ \citenamefont {Dubowik}}]{kudryavtsev}%
  \BibitemOpen
  \bibfield  {author} {\bibinfo {author} {\bibfnamefont {Y.}~\bibnamefont
  {Kudryavtsev}}, \bibinfo {author} {\bibfnamefont {N.}~\bibnamefont {Uvarov}},
  \bibinfo {author} {\bibfnamefont {V.}~\bibnamefont {Iermolenko}}, \bibinfo
  {author} {\bibfnamefont {I.}~\bibnamefont {Glavatskyy}}, \ and\ \bibinfo
  {author} {\bibfnamefont {J.}~\bibnamefont {Dubowik}},\ }\href@noop {}
  {\bibfield  {journal} {\bibinfo  {journal} {Acta Materialia}\ }\textbf
  {\bibinfo {volume} {60}},\ \bibinfo {pages} {4780} (\bibinfo {year}
  {2012})}\BibitemShut {NoStop}%
\bibitem [{\citenamefont {Gasi}\ \emph {et~al.}(2013)\citenamefont {Gasi},
  \citenamefont {Nayak}, \citenamefont {Nicklas},\ and\ \citenamefont
  {Felser}}]{Gasi}%
  \BibitemOpen
  \bibfield  {author} {\bibinfo {author} {\bibfnamefont {T.}~\bibnamefont
  {Gasi}}, \bibinfo {author} {\bibfnamefont {A.~K.}\ \bibnamefont {Nayak}},
  \bibinfo {author} {\bibfnamefont {M.}~\bibnamefont {Nicklas}}, \ and\
  \bibinfo {author} {\bibfnamefont {C.}~\bibnamefont {Felser}},\ }\href@noop {}
  {\bibfield  {journal} {\bibinfo  {journal} {Journal of Applied Physics}\
  }\textbf {\bibinfo {volume} {113}},\ \bibinfo {eid} {17} (\bibinfo {year}
  {2013})}\BibitemShut {NoStop}%
\bibitem [{\citenamefont {Caraballo~Vivas}\ \emph {et~al.}(2014)\citenamefont
  {Caraballo~Vivas}, \citenamefont {Rocco}, \citenamefont {Costa~Soares},
  \citenamefont {Caldeira}, \citenamefont {Coelho},\ and\ \citenamefont
  {Reis}}]{Caraballo}%
  \BibitemOpen
  \bibfield  {author} {\bibinfo {author} {\bibfnamefont {R.~J.}\ \bibnamefont
  {Caraballo~Vivas}}, \bibinfo {author} {\bibfnamefont {D.~L.}\ \bibnamefont
  {Rocco}}, \bibinfo {author} {\bibfnamefont {T.}~\bibnamefont {Costa~Soares}},
  \bibinfo {author} {\bibfnamefont {L.}~\bibnamefont {Caldeira}}, \bibinfo
  {author} {\bibfnamefont {A.~A.}\ \bibnamefont {Coelho}}, \ and\ \bibinfo
  {author} {\bibfnamefont {M.~S.}\ \bibnamefont {Reis}},\ }\href {\doibase
  http://dx.doi.org/10.1063/1.4892677} {\bibfield  {journal} {\bibinfo
  {journal} {Journal of Applied Physics}\ }\textbf {\bibinfo {volume} {116}},\
  \bibinfo {eid} {063907} (\bibinfo {year} {2014})}\BibitemShut {NoStop}%
\bibitem [{\citenamefont {Gschneidner~Jr}\ and\ \citenamefont
  {Pecharsky}(2000)}]{gschneidner2000}%
  \BibitemOpen
  \bibfield  {author} {\bibinfo {author} {\bibfnamefont {K.}~\bibnamefont
  {Gschneidner~Jr}}\ and\ \bibinfo {author} {\bibfnamefont {V.}~\bibnamefont
  {Pecharsky}},\ }\href@noop {} {\bibfield  {journal} {\bibinfo  {journal}
  {Annual Review of Materials Science}\ }\textbf {\bibinfo {volume} {30}},\
  \bibinfo {pages} {387} (\bibinfo {year} {2000})}\BibitemShut {NoStop}%
\bibitem [{\citenamefont {Carvalho}\ \emph {et~al.}(2014)\citenamefont
  {Carvalho}, \citenamefont {Tedesco}, \citenamefont {Pires}, \citenamefont
  {Soffner}, \citenamefont {Guimar\~{a}es}, \citenamefont {Mansanares},\ and\
  \citenamefont {Coelho}}]{Magnus2013}%
  \BibitemOpen
  \bibfield  {author} {\bibinfo {author} {\bibfnamefont {A.~M.~G.}\
  \bibnamefont {Carvalho}}, \bibinfo {author} {\bibfnamefont {J.~C.~G.}\
  \bibnamefont {Tedesco}}, \bibinfo {author} {\bibfnamefont {M.~J.~M.}\
  \bibnamefont {Pires}}, \bibinfo {author} {\bibfnamefont {M.~E.}\ \bibnamefont
  {Soffner}}, \bibinfo {author} {\bibfnamefont {A.~O.}\ \bibnamefont
  {Guimar\~{a}es}}, \bibinfo {author} {\bibfnamefont {A.~M.}\ \bibnamefont
  {Mansanares}}, \ and\ \bibinfo {author} {\bibfnamefont {A.~A.}\ \bibnamefont
  {Coelho}},\ }\href {\doibase http://dx.doi.org/10.1063/1.4861906} {\bibfield
  {journal} {\bibinfo  {journal} {Applied Physics Letters}\ }\textbf {\bibinfo
  {volume} {104}},\ \bibinfo {eid} {029902} (\bibinfo {year}
  {2014})}\BibitemShut {NoStop}%
\bibitem [{\citenamefont {Zhang}\ \emph {et~al.}(2003)\citenamefont {Zhang},
  \citenamefont {Bruck}, \citenamefont {Tegus}, \citenamefont {Buschow},\ and\
  \citenamefont {de~Boer}}]{Zhang2}%
  \BibitemOpen
  \bibfield  {author} {\bibinfo {author} {\bibfnamefont {L.}~\bibnamefont
  {Zhang}}, \bibinfo {author} {\bibfnamefont {E.}~\bibnamefont {Bruck}},
  \bibinfo {author} {\bibfnamefont {O.}~\bibnamefont {Tegus}}, \bibinfo
  {author} {\bibfnamefont {K.~J.}\ \bibnamefont {Buschow}}, \ and\ \bibinfo
  {author} {\bibfnamefont {F.}~\bibnamefont {de~Boer}},\ }\href {\doibase
  http://dx.doi.org/10.1016/S0921-4526(02)01853-7} {\bibfield  {journal}
  {\bibinfo  {journal} {Physica B: Condensed Matter}\ }\textbf {\bibinfo
  {volume} {328}},\ \bibinfo {pages} {295 } (\bibinfo {year}
  {2003})}\BibitemShut {NoStop}%
\end{thebibliography}
\end{document}